\documentclass[english,letter,12pt,fleqn]{article}

\usepackage{ulem} 
\usepackage{babel}
\usepackage{amsmath,amssymb,amsfonts,hhline}
\usepackage{graphicx}
\usepackage[hmargin=2.5cm,vmargin=2cm]{geometry}
\usepackage{natbib}

\usepackage{mathrsfs}

\usepackage{dpfloat}

\allowdisplaybreaks[1]

\newcommand\independent{\protect\mathpalette{\protect\independenT}{\perp}}
\def\independenT#1#2{\mathrel{\rlap{$#1#2$}\mkern2mu{#1#2}}}

\newcommand{\vectornorm}[1]{\left|\left|#1\right|\right|}

\makeatletter

\newcommand{\Rmnum}[1]{\expandafter\@slowromancap\romannumeral #1@}
\makeatother

\newtheorem{lemma}{Lemma}[section]

\newtheorem{definition}{Definition}[section]

\newtheorem{proposition}{Proposition}[section]

\bibpunct{(}{)}{;}{a}{}{,}

\numberwithin{equation}{section}

\begin{document}

\title{\huge{\bf Averaging orthogonal projectors}}
\author{ Eero Liski, Klaus Nordhausen, Hannu Oja, Anne Ruiz-Gazen \footnote{Eero Liski is a Researcher, School of Health Sciences, 33014 University of Tampere, Finland (email: eero.liski@uta.fi). Klaus Nordhausen is a University Lecturer, School of Information Sciences, 33014 University of Tampere, Finland and School of Health Sciences, 33014 University of Tampere, Finland (email: klaus.nordhausen@uta.fi). Hannu Oja is an Academy Professor, School of Health Sciences, 33014 University of Tampere, Finland (email: hannu.oja@uta.fi).
Anne Ruiz-Gazen is a Professor, Toulouse School of Economics, France (email: anne.ruiz-gazen@tse-fr.eu)} \\ }

\date{}
\maketitle

\begin{abstract}

Dimensionality is a major concern in analyzing large data sets. Some well known dimension reduction methods are for example principal component analysis (PCA), invariant coordinate selection (ICS), sliced inverse regression (SIR), sliced average variance estimate (SAVE), principal hessian directions (PHD) and inverse regression estimator (IRE). However, these methods are usually adequate of finding only certain types of structures or dependencies within the data. This calls the need to combine information coming from several different dimension reduction methods. We propose a generalization of the Crone and Crosby distance, a weighted distance that allows to combine subspaces of different dimensions. Some natural choices of weights are considered in detail. Based on the weighted distance metric we  discuss the concept of averages of subspaces as well to combine various dimension reduction methods. The performance of the weighted distances and the combining approach is illustrated via simulations.


\newline

  \end{abstract}

\section{Introduction}

Dimension reduction plays an important role in high dimensional data analysis. One then wishes to reduce the dimension of a $p$-variate random vector $x = (x_1,\dots, x_p)'$  using a  transformation  $z = B'x$  where the transformation matrix $B$  is a $p \times k$ matrix  with linearly independent columns, $k\le p$.
The column vectors of $B$ then span the $k$-dimensional subspace of interest. The transformation to the subspace can also be done using the corresponding $p \times p$ orthogonal projector $P_B = B(BB')^{-1}B'$. The transformation $z = P_Bx$ projects the observations to a linear $k$-variate subspace.

There are two main types of dimension reduction -- unsupervised and supervised. Principal component analysis (PCA) is perhaps the best known unsupervised dimension reduction method. PCA finds an orthogonal transformation matrix in such a way that the components in the new coordinate system are uncorrelated and ordered according to their variances. In dimension reduction, only the $k$ components with highest variances are taken. Independent component analysis (ICA) is another example of an unsupervised dimension reduction method. The fourth-order blind identification (FOBI) \citep[][]{Cardoso:1989} procedure then finds a transformation matrix in such a way that the new components are uncorrelated with respect to two distinct scatter matrices, the regular covariance matrix and a scatter matrix based on fourth moments, and ordered according to their kurtosis values. Then the $k$ components with kurtosis values most deviating from that of a normal distribution are most interesting.

In dimension reduction, the goal is often to use the transformed (reduced) variables to predict the value of a known response variable $y$. In supervised dimension reduction, the joint distribution of $x$ and $y$ is then taken into consideration in the dimension reduction of $x$, and it is hoped that
 $y\independent x|B'x$.
 Sliced inverse regression (SIR) \citep[][]{Li:1991} is a well known supervised dimension reduction method.  It is, along with ICA, based on the comparison of two scatter matrices, where the second scatter matrix depends - unlike in unsupervised dimension reduction methods - on the joint distribution of $x$ and $y$. Other well known dimension reduction methods are for example the sliced average variance estimate (SAVE) and the principal hessian directions (PHD) \citep[see][respectively]{CookWeisberg:1991, Li:1992}.  A recent contribution to the field of supervised dimension reduction methods is the supervised invariant coordinate selection (SICS) \citep[][]{Liski:2012}. It is an extension of invariant coordinate  selection (ICS) \citep[][]{Tyler:2009}. Both ICS and SICS are based on the comparison of two scatter matrices $S_1$ and $S_2$, and the transformation matrix is based on the eigenvalues and eigenvectors  of $S_1^{-1}S_2$. In SICS however, the second scatter matrix depends on the joint distribution of $x$ and $y$.

Individual dimension reduction methods are usually adequate to find only special types of subspaces or special relationships between $x$ and $y$. For example, SIR works well for linear relationships but may completely fail for other type of dependencies \citep[see for example][]{CookWeisberg:1991}. Hence, there is a need for new approaches, which would use the good properties of individual dimension reduction methods and combine the information in an efficient way.  The aim of this paper is to combine different dimension reduction methods in such a way that the ``best qualities" of each method are picked up.  To combine different dimension reduction methods is to say that we combine the individual orthogonal projectors possibly with various ranks and find an ``average orthogonal projector" (AOP)  with an optimized rank. Our approach is similar to  the approach in \cite{Crone:1995}. The idea is to find the AOP which is, on the average, closest to individual orthogonal projectors with respect to some distance criterium.

The paper is organised as follows. In Section~\ref{sec: theory} we discuss subspaces and propose a generalization of the Crone and Crosby distance. \cite{Crone:1995} considered subspaces of equal dimensions, whereas our weighted distance allows subspaces of different dimensions. Some natural choices of weights are given. Furthermore, the concept of averages of subspaces is discussed. In Section~\ref{sec: applications} the performance of the weighted distance and the different AOP's is evaluated in two unsupervised dimension reduction applications and one supervised dimension reduction simulation study. The paper ends with some final remarks.

\section{ Subspaces and distances between subspaces}\label{sec: theory}

\subsection{Subspaces with the same dimension $k$}

We first consider linear subspaces in $\mathbb{R}^p$ with a fixed dimension $k$, $1\le k < p$. A linear subspace and the distances between the subspaces can be defined in several ways.
\begin{enumerate}
\item
The subspace is defined as a  linear subspace spanned by the linearly independent columns of a $p\times k$ matrix $B$, that is,
$
\mathcal{S}_B=\{ B a\ : a\in \mathbb{R}^k  \}.
$
This definition based on a matrix $B$ is a bit obscure in the sense that $\mathcal{S}_B=\mathcal{S}_{BA}$ for all full-rank $k\times k$ matrices $A$.
According to this definition, the same subspace can in fact be fixed by any member in a set of matrices equivalent to $B$,
$$ \{ BA\ : \ \mbox{A is a full-rank $k\times k $ matrix}\}. $$
The non-uniqueness of $B$ may cause technical problems in the estimation of a subspace. Consider two $p \times k$ matrices $B_1$ and $B_2$ with rank $k$. Then  a measure of distance between subspaces spanned by $B_1$ and $B_2$ can be defined as
$k-\sum_{i=1}^k \rho_i^2=k-tr(R'R)$
where $\rho_1^2,\ldots,\rho_k^2$ are the squared canonical correlations between $B_1$ and $B_2$ \citep[][]{Hotelling:1936} and $R=(B_1'B_1)^{-1/2}B_1'B_2(B_2'B_2)^{-1/2}$.
Note that if $B_1$ and $B_2$ are equivalent then the squared canonical correlations are all 1.
\item
The subspace is defined as a  linear subspace spanned by the orthonormal columns of a $p\times k$ matrix $ U$.  Note that, starting with $B$, one can choose
$U=B(B'B)^{-1/2}$ for this second definition. Unfortunately, the definition is still obscure as $\mathcal{S}_U=\mathcal{S}_{UV}$ for all orthonormal $k\times k$ matrices $V$, and the same subspace is given by any matrix in the class of equivalent orthonormal matrices
$$ \{ UV\ : \ \mbox{V is an orthonormal $k\times k $ matrix}\}. $$
The principal angles $\theta_i\in[0,\pi/2]$ between the subspaces  $U_1$ and $U_2$ with corresponding $k$-variate direction vectors $u_i$ and $v_i$ $i=1,\ldots,k$,   are recursively defined by maximizing
$
u_i'(U_1'U_2)v_i
$
subject to the constraints
$u_i'u_i=v_i'v_i=1$, and $u_i'u_j=v_i'v_j=0$, $j=1,\ldots,i-1$.
The $i$th principal angle is then
$\cos \theta_i= u_i'(U_1'U_2)v_i$, $i=1,...,k$,
and a measure of distance between the subspaces may be obtained as  $k-\sum_{i=1}^k \cos^2 \theta_i=k-\sum_{i=1}^k (u_i'v_i)^2 $. It is easy to see that it equals  to $k-\sum_{i=1}^k \rho_i^2$.
\item The subspace is defined as the linear subspace given by an orthogonal projector $P$, that is, a $p\times p$ transformation matrix $P$ such that
\[
(x_1-Px_1) \perp Px_2\  \mbox{for all}\ x_1,x_2\in \mathbb{R}^p\ \  \mbox{which is equivalent to} \ \  P=P'=P^2.
\]
Matrix $P$ provides a unique way to fix the subspace $\mathcal S_P=\{Px : x\in \mathbb{R}^p \}$. Note that, starting from $B$, one can define $P=P_B=B(B'B)^{-1}B'$ in a unique way. Starting from $U$ gives similarly $P=P_U=UU'$. The squared distance between the subspaces given by two orthogonal projectors $P_1$ and $P_2$ may then be  defined as the matrix (Frobenius) norm
$$||P_1-P_2||^2=2(k-tr(P_1P_2))=2(k-\sum_{i=1}^k \cos^2 \theta_i)=2(k-\sum_{i=1}^k \rho_i^2).$$
\cite{Crone:1995} use
\[
D(P_1, P_2) =  \left(k - tr(P_1 P_2)\right)^{1/2} = \frac 1{\sqrt{2}}\vectornorm{P_1 - P_2}
\]
as a distance between two $k$-dimensional subspaces of $\mathbb R^p$ given by orthogonal projectors $P_1$ and $P_2$.
It is then easy to see that
$
0\le D^2(P_1, P_2) \le min\{k,p-k\}
$
and that the distance obeys the triangular inequality
$
D(P_1,P_3)\le D(P_1,P_2)+D(P_2,P_3)
$
for any orthogonal projectors $P_1$, $P_2$ and $P_3$.
\end{enumerate}

\subsection{Subspaces with arbitrary dimensions}

Assume next that the ranks of the projection matrices $P_1$ and $P_2$ are $k_1$ and $k_2$, respectively, where $k_1,k_2=0,...,p$. For completeness of the theory, we also accept projection matrices $P=0$ with rank $k=0$. As
$||P_1-P_2||^2\ge |k_1-k_2|$, one
possible extension of the above distance is
$
D(P_1, P_2) =  \frac 1{\sqrt{2}}\left[ \vectornorm{P_1 - P_2}^2-|k_1-k_2| \right]^{1/2}.
$
Then
$
0\le D^2(P_1, P_2) \le min\{k_1,k_2,p-k_1,p-k_2\}
$
but, unfortunately, the triangular inequality is not true for this distance.
We therefore consider other  extensions of  the metric by \cite{Crone:1995}.

Let $w(0),\ldots,w(p)$ be positive weights attached to dimensions $0,\ldots,p$. (We will later see that the choice of $w(0)$ is irrelevant for the theory.)
 We then give the following definition.

\begin{definition}
A weighted distance between subspaces  $P_1$ and $P_2$ with ranks $k_1$ and $k_2$ is given by
\begin{equation}\label{formula: genform}
D_{w}^2(P_1,P_2)=\frac 12 \vectornorm{w(k_1)P_1-w(k_2)P_2}^2.
\end{equation}
\end{definition}
The weights are used  to make the orthogonal projectors $P_1$ and $P_2$ with different ranks more comparable in some sense.
As the distance $D_{w}(P_1,P_2)$ is based on  the matrix (Frobenius) norm, (i) $D_{w}(P_1,P_2)\ge 0$,
(ii)  $D_{w}(P_1,P_2)=0$ if and only if $P_1=P_2$, (iii)  $D_{w}(P_1,P_2)=D_{w}(P_2,P_1)$, and
(iv) $D_{w}(P_1,P_3)\le D_{w}(P_1,P_2)+D_{w}(P_2,P_3)$, and  we have the following result.

\begin{proposition}\label{Metricwithweights}
For all weight functions $w$,  $D_{w}(P_1,P_2)$ is a metric in the space of orthogonal projectors, and the
strict lower and upper bounds of $D_{w}^2(P_1,P_2)$ for  the dimensions $k_1$ and $k_2$ are
\[
m(k_1,k_2) - w(k_1)w(k_2) \min\{k_1,k_2 \} \le
D_{w}^2(P_1,P_2) \le
m(k_1,k_2) + w(k_1)w(k_2) \min\{p-k_1-k_2,0 \}
\]
where
\[
m(k_1,k_2)=\frac {w^2(k_1)k_1+w^2(k_2)k_2} 2.
\]
\end{proposition}

Some interesting choices of the weights are, for $k>0$,
\[
(a)\ w_a(k)= 1, \ \ \ (b)\ w_b(k)=\frac 1k,\ \ \mbox{and}\ \ (c)\  w_c(k)=\frac 1{\sqrt{k}}.
\]
Weights in ($a$) give the distance by \cite{Crone:1995}. Weights in ($b$) and ($c$) standardize the matrices so that
$tr(w(k_i)P_i)=1$ and $||w(k_i)P_i||=1$, respectively, if $k_i>0$.
It is  remarkable that
\[
D_{w_c}^2(P_1,P_2)=1-\frac{tr(P_1 P_2)}{\sqrt{tr(P_1) tr(P_2)}}
\]
where
\[
\frac{tr(P_1 P_2)}{\sqrt{tr(P_1) tr(P_2)}}=\frac{vec(P_1)'vec(P_2)}{ \sqrt{vec(P_1)'vec(P_1)} \sqrt{vec(P_2)'vec(P_2)}}
\]
is a correlation between vectorized $P_1$ and $P_2$.

Proposition~\ref{Metricwithweights} implies that, for nonzero $k_1$ and $k_2$,
 the distances $D_{w}^2(P_1,P_2)$  get any values on the closed intervals
\begin{eqnarray*}
   &(a):& \left[\frac 12 |k_1-k_2|, \frac 12 (k_1+k_2)+\min\{ p-k_1-k_2,0\}\right],  \\
  &(b):&  \left[\frac 12 \big|k_1^{-1}-{k_2}^{-1}\big|, \frac 12\left({k_1}^{-1}+{k_2}^{-1} \right)+{k_1^{-1}k_2^{-1}} \min\{ p-k_1-k_2,0\}\right],\ \ \mbox{and}\\
  &(c):& \left[ 1-\min\{k_1^{1/2}k_2^{-1/2},k_1^{-1/2}k_2^{1/2}  \}, 1+ k_1^{-1/2} k_2^{-1/2}\min\{ p-k_1-k_2,0\}\right].
\end{eqnarray*}
If $k_1=0$, for example, then $D_{w}^2(P_1,P_2)$ is simply $w^2(k_2)k_2/2$.
Recall that, for all three choices of weights, the distance is zero only if $P_1=P_2$ (and $k_1=k_2$).
For weights $w_a$, the largest possible value for $D_w^2(P_1,P_2)$ is $p/2$ and it is obtained if and only if $P_1$ and $P_2$ are orthogonal  and $P_1+P_2=I_p$
(i.e., $k_1+k_2=p$). For weights $w_b$, $D_w^2(P_1,P_2)\le 1$,  and $D_w^2(P_1,P_2)=1$ if and only if $P_1$ and $P_2$ are orthogonal and $k_1=k_2=1$. Finally, for weights $w_c$, the maximum value $D_w(P_1,P_2)=1$ for $k_1,k_2\ne 0$ is attained as soon as $P_1$ and $P_2$ are orthogonal and $k_1+k_2\le p$.

The following two special cases illustrate the differences between the three distances.
\begin{enumerate}
\item
First, consider the case when $\mathcal{S}_{P_1}\subset \mathcal{S}_{P_2}$. Then naturally $tr(P_1P_2)=tr(P_1)=k_1$ and
\[
D_{w}^2(P_1,P_2)=  \frac {w^2(k_1)k_1+w^2(k_2)k_2} 2 -w(k_1)w(k_2)k_1
\]
and therefore, for $k_2\ne 0$ and with $\lambda=k_1/k_2$,
\begin{eqnarray*}
D_{w_a}^2(P_1,P_2) &=& \frac {k_2}2 (1-\lambda), \\
D_{w_b}^2(P_1,P_2) &=& \frac 1{2k_1} (1-\lambda), \ \ \mbox{and} \\
D_{w_c}^2(P_1,P_2) &=& 1-\sqrt{\lambda}.
\end{eqnarray*}
One can see that $D_{w_c}^2(P_1,P_2)$ depends only on the ratio between $k_1$ and $k_2$, which can be seen as a nice feature. $D_{w_a}^2(P_1,P_2)$ and $D_{w_b}^2(P_1,P_2)$ however depend additionally on the actual values of $k_1$ and $k_2$.
\item
Second, consider the case when $\mathcal{S}_{P_1}$ and $\mathcal{S}_{P_2}$ are orthogonal, that is, when $tr(P_1P_2)=0$. Then
\[
D_{w}^2(P_1,P_2)=  \frac {w^2(k_1)k_1+w^2(k_2)k_2} 2
\]
and therefore, for nonzero $k_1$ and $k_2$,
\begin{eqnarray*}
D_{w_a}^2(P_1,P_2) &=& \frac 12 (k_1+k_2), \\
D_{w_b}^2(P_1,P_2) &=& \frac 12\left(\frac 1{k_1}+\frac 1{k_2} \right), \ \ \mbox{and} \\
D_{w_c}^2(P_1,P_2) &=& 1.
\end{eqnarray*}
It is natural to think subspaces that are orthogonal to each other are furthest apart possible. This information is apparent in $D_{w_c}^2(P_1,P_2)$. However, interpreting both $D_{w_a}^2(P_1,P_2)$ and $D_{w_b}^2(P_1,P_2)$ is again more difficult since they depend on the actual values of $k_1$ and $k_2$.

\end{enumerate}

\subsection{Averages of subspaces with arbitrary dimensions}

Consider orthogonal projectors $P_1,\ldots,P_m$ with ranks $k_1,\ldots,k_m$. To combine the projectors we give the following

\begin{definition}\label{avedef}
The average orthogonal projector (AOP)  $\, P_w$ based on weights $w(0),\ldots,w(p)$ is an orthogonal projector that minimizes the objective function
\[
\sigma_w^2(P)= \frac 1m \sum_{i=1}^m D_w^2(P_i,P).
\]
\end{definition}

To find the AOP, we can use the following result.

\begin{lemma}\label{AOPlemma}
The AOP  $P_w$ maximizes the function
\[
D(P)= w(k)tr(\bar P_wP)- \frac 12 w^2(k)k,
\]
where
\[
\bar P_w=\frac 1m \sum_{i=1}^m w(k_i) P_i
\]
is a regular average of weighted projectors, and $k$ is the rank of $P$.
\end{lemma}

Naturally,  $\bar P_w$ is symmetric and non-negative definite, but not a projector anymore. In the following derivations, we need  its eigenvector and eigenvalue decomposition
\[
\bar P_w=U \Lambda U'=\sum_{i=1}^p \lambda_i u_i u_i'
\]
where  $\lambda_1\ge \ldots \ge \lambda_p\ge 0$ and $u_i$ is the eigenvector
corresponding to the eigenvalue $\lambda_i$. Recall that the eigenvectors are uniquely defined only for eigenvalues that are distinct from other eigenvalues.
Using the Lemma~\ref{AOPlemma} and the eigenvector and eigenvalue decomposition $\bar P_w$, our main result easily follows.

\begin{proposition}
The rank $k$ of the AOP $\, P_w$ maximizes the function
\[
f_w(k)=w(k)(\sum_{i=1}^k \lambda_i)I(k>0) -\frac 12 w^2(k) k,\ \ k=0,\ldots,p,
\]
where $\lambda_1\ge \ldots\ge \lambda_p\ge 0$ are the eigenvalues of $\bar P_w$. Moreover,
\[
P_w=I(k>0)\cdot \sum_{i=1}^k u_i u_i'
\]
where $u_1,\ldots,u_k$ are the eigenvalues corresponding to eigenvalues $\lambda_1,\ldots,\lambda_k$.
\end{proposition}

Note that the calculation of the AOP $P_w$ is easy, only the eigenvalues and eigenvectors of $\bar P_w$ are needed. The AOP $P_w$ is not always unique. This happens for example if  the rank  of an AOP is $k$ and  $\lambda_{k+1}=\lambda_{k}$.
Consider now the three weight functions
\[
(a)\ w_a(k)= 1, \ \ \ (b)\ w_b(k)=\frac 1k,\ \ \mbox{and}\ \ (c)\  w_c(k)=\frac 1{\sqrt{k}}, \ \  \mbox{for $k>0$}.
\]
The function $f_w$ for these three weight functions is, for $k>0$,
\[
   (a) \  \sum_{i=1}^k \lambda_i -\frac k2,  \quad
   (b) \  \frac 1k \left(\sum_{i=1}^k \lambda_i -\frac 12\right)   ,\ \ \mbox{and}\ \
   (c)  \ \frac 1{\sqrt{k}} \left(\sum_{i=1}^k \lambda_i\right) -\frac 12 .
\]

Note  that  $f_w(0)=0$ for all weights $w$. To find local maxima for these functions, one can then use the results
\begin{eqnarray*}
   &(a):& f_w(k+1)\ge f_w(k) \ \ \Leftrightarrow \ \ \lambda_{k+1}\ge \frac 12,  \\
  &(b):&   f_w(k+1)\ge f_w(k) \ \ \Leftrightarrow \ \ \lambda_{k+1}\ge \frac 1k \left(\lambda_1+\ldots+\lambda_k- \frac 1{2}\right)  ,\ \ \mbox{and}\\
  &(c):&  f_w(k+1)\ge f_w(k) \ \ \Leftrightarrow \ \ \lambda_{k+1}\ge \left(\sqrt{\frac {k+1}k}-1\right) (\lambda_1+\ldots+\lambda_k)
\end{eqnarray*}
for $k=1,\ldots,p-1$.

Note that for $(a)$, $f_w(k)$ is a concave function and the global maximum is simply the largest $k$ with the eigenvalue $\lambda_k\ge \frac 12$.  The functions in $(b)$ and $(c)$ are not concave, however, and the global maximum is found by computing all the values $f_w(k)$, $k=0,...,p$.

\section{Applications}\label{sec: applications}
In this section we discuss the performance of the averages of orthogonal projectors (AOP) for three different dimension reduction problems.
 The orthogonal  projectors and their combinations aim for different targets in different applications.  Each problem with natural projectors will
be first shortly introduced, and then the performance of AOP is demonstrated using simulation studies. The computations in this section are done using R \citep{R:2012} by mainly using the packages dr \citep{Weisberg:2002}, MNM \citep{NordhausenOja:2011}, pcaPP \citep{pcaPP:2012} and robustbase \citep{robustbase:2012}.

\subsection{Principal component Analysis}

Classical principal component analysis (PCA) may be based  on the eigenvector and eigenvalue decomposition of the covariance matrix of a $p$-variate random vector $x$, that is, on
\[
cov(x)= U\Lambda U'=\sum_{i=1}^p \lambda_i u_i u_i'
\]
where $\lambda_1\ge ...\lambda_p\ge 0$ are the ordered eigenvalues and $u_1,...,u_p$ are the corresponding eigenvectors. Orthogonal projector
$P_{cov}=\sum_{i=1}^k  u_i u_i'$
then projects $p$-variate observations to the $k$-variate subset with maximum variation. It is unique if $\lambda_{k+1}>\lambda_k$.

Let $F_x$ be the cumulative distribution function of $x$. A $p\times p$ matrix valued functional $S(F)$ is a scatter matrix if $S(F)$ is a non-negative definite and symmetric matrix with the affine equivariance property
\[
S(F_{Ax+b})=AS(F_x)A'\ \ \mbox{for all full-rank $p\times p$ matrices $A$ and all $p$-vectors $b$.}
\]
It is remarkable that, if $x$ has an elliptic distribution then the ordered eigenvectors of $S(F_x)$ are those of $cov(x)$. Therefore, in the elliptic case, any scatter matrix can be used to find $P=\sum_{i=1}^k  u_i u_i'$ and the matrix $P$ is  a well-defined population quantity  even if the second moments (and the covariance matrix)  do not exist. Naturally, the sample statistics corresponding to different scatter matrices then have different statistical (efficiency and robustness) properties. For a fixed value of $k$, one can then try to  ``average'' these different PCA approaches to get a compromise estimate.

We next illustrate the performance of the AOP in the following simple scenario. Let first $x\sim N_6(0,\Lambda)$
$\Lambda = diag(9,7,5,1,1,1)$.  We choose $k=3$ and wish to estimate $P_{cov}=diag(1,1,1,0,0,0)$.
Let then $x_1,...,x_n$ be a random sample from $x\sim N_6(0,\Lambda)$, and find an estimate  $P_{\widehat{cov}}$, an orthogonal projector with rank $k=3$ obtained from the sample covariance matrix. This estimate is then
combined with three robust estimates, namely,
\begin{description}
  \item[$P_{Tyler}$] that is based on Tyler's shape matrix \citep{Tyler:1987}  with the affine equivariant version of spatial median as a multivariate location estimate \citep{HettmanspergerRandles:2002}.
  \item[$P_{MCD}$] that is based on the minimum covariance determinant (MCD) estimator \citep{Rousseeuw:1986}.
  \item[$P_{PP}$] that is based on  projection pursuit (PP) approach for PCA with the median absolute deviation (mad) criterion as suggested in \citet{CrouxRuizGazen:2005}.
\end{description}
In the simulations, $x_1,...,x_n$ was a random sample from $N_6(0,\Lambda)$ with $n=400$,  and the sampling was repeated 1000 times.
As $k_1=...=k_m=k=3$ is fixed ,
we use only $w_a$ as the weight function.
The average squared distances $D^2_{w_a}$ between the four projector estimates, their AOP $P_{w_a}$, and $P_{cov}$ are shown in Table~\ref{tab: distmatrix}. A similar simulation study was conducted but with observations coming from a heavy-tailed elliptical $t_2$ distribution with  $9,7,5,1,1,1$ as proportional eigenvalues. Note that the regular scatter matrix does not exist in this case but the true projection matrix is still well defined.

\begin{table}[htbp]
\centering
\caption{Average squared distances $D_{w_a}^2$ between the four projector estimates, their AOP $P_{w_a}$, and true $P_{cov}$. For all projectors, rank $k=3$. The averages are based on 1000 random samples of size $n=400$ from
 $N_6(0,\Lambda)$.}%
\label{tab: distmatrix}
\begin{tabular}{lllllll}
    \hline
       & $P_{\widehat{cov}}$ & $P_{Tyler}$ & $P_{MCD}$ & $P_{PP}$ & $P_{w_a}$ & $P_{cov}$ \\
    \hline
    $P_{\widehat{cov}}$    & 0.000 & 0.002  & 0.002 & 0.056 & 0.004 & 0.005 \\
    $P_{Tyler}$  & 0.002 & 0.000  & 0.001 & 0.054 & 0.004 & 0.007\\
    $P_{MCD}$    & 0.002 & 0.001  & 0.000 & 0.055 & 0.004 & 0.007\\
    $P_{PP}$     & 0.056 & 0.054  & 0.055 & 0.000 & 0.031 & 0.061\\
    $P_{w_a}$   & 0.004 & 0.004  & 0.004 & 0.031 & 0.000 & 0.009\\
    \hline
\end{tabular}
\end{table}

The results in the multivariate normal case show, as expected,  that the projector estimate based on the covariance matrix is the best one here.
Also the average projector performs very well although it combines information coming from much worse $P_{PP}$. In the $t_2$ distribution case,
traditional $P_{\widehat{cov}}$ fails but the average projector is still performing well (see Table~\ref{tab: distmatrix_t2}). Recall the second moments and  $P_{cov}$ do not exist in this case.


\begin{table}[htbp]
\centering
\caption{Average squared distances $D_{w_a}^2$ between the four projector estimates, their AOP $P_{w_a}$, and true $P_{cov}$. For all projectors, rank $k=3$. The averages are based on 1000 random samples of size $n=400$ from
 $t_2$ with eigenvalues $9,7,5,1,1,1$.}%
\label{tab: distmatrix_t2}
\begin{tabular}{lllllll}
    \hline
       & $P_{\widehat{cov}}$ & $P_{Tyler}$ & $P_{MCD}$ & $P_{PP}$ & $P_{w_a}$  & $P_{cov}$ \\
    \hline
    $P_{\widehat{cov}}$    & 0.000 & 0.110 & 0.122 & 0.169 & 0.074 & 0.114 \\
    $P_{Tyler}$  & 0.110 & 0.000 & 0.006 & 0.063 & 0.010 & 0.007\\
    $P_{MCD}$    & 0.122 & 0.006 & 0.000 & 0.066 & 0.014 & 0.012\\
    $P_{PP}$     & 0.169 & 0.063 & 0.066 & 0.000 & 0.042 & 0.070\\
    $P_{w_a}$  & 0.074  & 0.010  & 0.014 & 0.042 & 0.000 & 0.016\\
    \hline
\end{tabular}
\end{table}

\subsection{Averaging one-dimensional PP projectors}

In the previous section we used projection pursuit (PP) approach for principal component analysis. PP is a much more general technique, however, and there are many other types of indices than just measures of variation to define ``interesting'' one-dimensional directions. PP actually dates back to \cite{Friedman:1974} and usually one searches for nongaussian directions. For a recent review of the existing indices, see for example \cite{RodriguezMartinezGoulermasMuRalph:2010}. A major challenge in PP is that it is computationally difficult to find the direction which globally maximizes the index and that there are usually several local maxima. However, since the local maxima may be also of interest, one possible strategy, as detailed in \citet{Ruiz-Gazen:2010}, is to run the algorithm many times using different initializations.
With this strategy, the user has many projectors of rank one but many of them are usually redundant. So, it is of particular interest to  summarize all these projectors in order to extract the directions that are useful and unique. It means that, in that case, one is interested in an average projector of projectors with rank one that may have a higher rank.

To demonstrate the interest of AOP in the context of PP, we choose the deflation-based fastICA method \citep{Hyvarinen:1999} as an example since it is well-understood and computationally quite simple. While deflation-based fastICA is originally developed in the context of independent component analysis (ICA), it can be seen as a traditional PP approach when only one direction is extracted.
For a random variable $x$ with the standardized version $z= cov(x)^{-1/2}(x-E(x))$, deflation-based fastICA maximizes a measure of non-gaussianity of the form $|E(G(u'z))|$, under the constraint that $u'u=1$, where $G$ is a selected twice differentiable nonlinear nonquadratic function with $G(0)=0$. The final PP direction is then $(u'cov(x)^{-1} u)^{-1/2} cov(x)^{-1/2}u$, and the corresponding orthogonal projector is $(u'cov(x)^{-1} u)^{-1}cov(x)^{-1/2}uu'cov(x)^{-1/2}$. In our simulations, we use four common choices of $G(u)$ with derivative functions $g(u)$:
(i) $u^3$ , (ii) $\tanh(u)$,  (iii) $u \exp(u^2 /2)$, and (iv) $u^2$. If there are more than one non-gaussian direction in the data, the direction to be found depends heavily on the initial value of the algorithm, see e.g. \cite{NordhausenIlmonenMandalOjaOllila:2011}.

In our simulation study, we choose a 10-variate $x=(x_1,\ldots,x_{10})'$ where the first three variables are mixtures of gaussian distributions and $x_i \sim N(0,1)$, for $i=4,\ldots,10$. More precisely, $x_1 =\frac{1}{\sqrt{5}}(p_1 y_1 + (1-p_1) y_2 - 2)$ with $p_1 \sim Bin(1,0.5)$, $y_1 \sim N(0,1)$ and $y_2 \sim N(4,1)$~;
$x_2 = \frac{1}{\sqrt{2.89}}(p_2 y_3 + (1-p_2) y_4 - 2.1)$ with $p_2 \sim Bin(1,0.3)$, $y_3 \sim N(0,1)$ and $y_4 \sim N(3,1)$, and $x_3 = \frac{1}{\sqrt{24.36}}(p_3 y_5 + (1-p_3) y_6 - 2)$ with $p_3 \sim Bin(1, 0.4)$, $y_5 \sim N(0,9)$ and $y_6 \sim N(8,9)$. We generated 1000 random samples of sizes $n=200$ from the 10-variate distribution described above. For each sample, we found 100 one-dimensional PP directions (4 choices of $G$, 25 random initial values for the algorithm for each choice of $G$). For each sample, 100 PP projectors were then averaged using each of the three weight functions $w_a$, $w_b$ and $w_c$. The average projector should in this setting then be close to the projector $P_{true} = diag(1,1,1,0,\ldots,0)$  with rank 3 that picks the three non-gaussian components of the data.
\begin{figure}[tbph]
    \centering
    \includegraphics[height = 5.5cm]{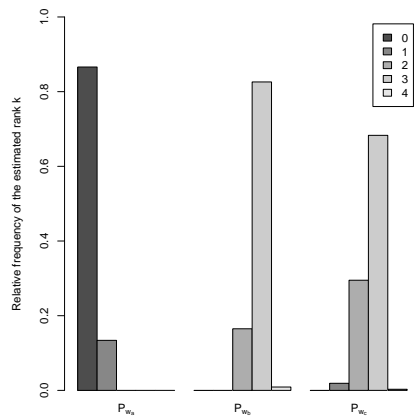}
    \caption{Relative frequencies of the estimated ranks of the AOP using the weight functions $w_a$, $w_b$ and $w_c$.}
    \label{fig: rank_freq} 
\end{figure}

Figure~\ref{fig: rank_freq} shows the relative frequencies of the  ranks of the AOPs obtained with the three weight functions in 1000 repetitions.
Clearly the weight function $w_a$ is not appropriate in this application because  $k_1=k_2=\ldots=k_m=1$ implies that $\sum_{i=1}^m\lambda_i=1$ with $\lambda_i\geq 0$, which means that there cannot be more than one eigenvalue larger than 1/2 and, consequently,  the rank $k$ equals zero or one. With weight functions $w_b$ and $w_c$,
the correct rank 3 is obtained in 82.6\% and 68.3\% of the runs, respectively.  It is also hoped that
the AOPs are close to the true projector $P_{true}$. To evaluate that, we found that the the average distances $D^2_{w_a}$
between ${P}_{w_a}$ and $P_{true}$, between ${P}_{w_b}$ and $P_{true}$, and between ${P}_{w_c}$ and $P_{true}$ were
1.005, 0.122, and  0.199, respectively. The same numbers for the distances based on $w_b$ and $w_c$ are
0.335, 0.018, and  0.035, and 0.425, 0.043, and 0.0736, respectively. Notice that, for all distances, the AOP  ${P}_{w_b}$ is closest on average to the true value
$P_{true}$.



\subsection{Supervised dimension reduction}
In the PCA application, we used the same  $k=3$ for orthogonal projectors and their AOP. In  the PP application, the rank of the orthogonal projectors  was taken as one
while the rank of their AOP was not fixed. However, for many dimension reduction methods, the ranks of the individual orthogonal projectors are not fixed but also estimated from the data, and the ranks may differ from one method to another. We now {look at} this scenario in the framework of supervised dimension reduction.

In supervised dimension reduction,  one often assumes that  a response variable  $y$  and the $p$-vector $x$  are related through
\[
y=f(b_1^Tx, \ldots, b_k^Tx, \epsilon),
\]
with an unknown function $f$ and an unknown error variable $\epsilon$. The goal of supervised dimension reduction is to estimate  the value of $k$ and the
matrix $B=(b_1, \ldots, b_k)$ to obtain $P_B$ with rank $k$. Hence, for supervised dimension reduction,  the joint distribution of $y$ and $x$ is of interest and, for the matrix $B$, it holds that $y\independent x| B'x$.

Many supervised dimension reductions are suggested in the literature and their performances often strongly  depend  on the unknown function $f$. The well-known sliced inverse regression (SIR) for example may not find  directions with nonlinear dependencies while, on the other hand, principal Hessian directions (PHD) cannot find linear relationships. Hence, when using supervised dimension reduction methods in practice, the estimated rank $k$ and the corresponding projector might differ considerably depending on the method. We propose to use the AOP in order to summarize in an efficient way the information brought by the complementary estimation strategies.

In our example we generate data sets from the following three models.
\begin{eqnarray*}
\text{M1:} \quad &y =&  b_{11}'  x + ( b_{12}'  x)^2 + \sigma \varepsilon \\
\text{M2:} \quad &y =& 5 +  b_{21}'  x + \sigma \varepsilon \\
\text{M3:} \quad &y =& ( b_{31}'  x)^{2} + \sigma \varepsilon
\end{eqnarray*}
where $ x \sim N_{10}( 0, I_{10})$, $\varepsilon \sim N(0,1)$, $\sigma = 0.5$ and $b_{ij}$'s are all 10-dimensional row vectors
\begin{eqnarray*}
 b_{11} &=& (2,3,0,\dots,0)' \ \text{and} \  b_{12} = (0,0,-5,0,\dots,0)', \\
 b_{21} &=& (1,1,1,0,\dots,0)', \ \ \mbox{and} \\
 b_{31} &=& (1,0,0,0,\dots,0)'
\end{eqnarray*}
Hence $k=2$ for model M1 and $k=1$ for models M2-M3. In each case, we generated 100  samples of size 400.

In our illustration, we use supervised dimension reduction methods implemented in the dr package that provide both the estimate of $k$ and the orthogonal projector estimate with the same rank $k$.  The estimation strategies  are then (i) sliced inverse regression (SIR), (ii) sliced variance estimation (SAVE), (iii) inverse regression estimation (IRE), and three types of principal hessian directions (PHD), namely, (iv) response based principal hessian directions (PHDY), (v)
residual based principal hessian directions (PHDR), and (vi) the so called $q$-based principal hessian directions (PHDQ). For details about these estimation methods, see \citet{Weisberg:2002} and references therein. We also add here (vii) PCA  with  $k$ chosen simply as the number of eigenvalues larger than 1. Naturally, PCA ignores $y$ and is therefore not supervised. (Its use could be motivated by the aim to avoid directions with small variation. In our case it just provides random projectors.)

In the following we want to compare the seven methods above and their AOPs based on the weight functions $w_b$ and $w_c$. The use of $w_a$ is not reasonable with varying $k$. We consider here the following four AOPs.
\begin{enumerate}
\item[{AOP1}:] The AOP using $w_b$ with fixed and true $k$.
\item[{AOP2}:] The AOP using $w_c$ with fixed and true $k$.
\item[{AOP3}:] The AOP using $w_b$ with estimated $k$.
\item[{AOP4}:] The AOP using $w_c$ with estimated $k$.
\end{enumerate}

\begin{figure}[tbph]
    \centering
    \includegraphics[height = 5.5cm]{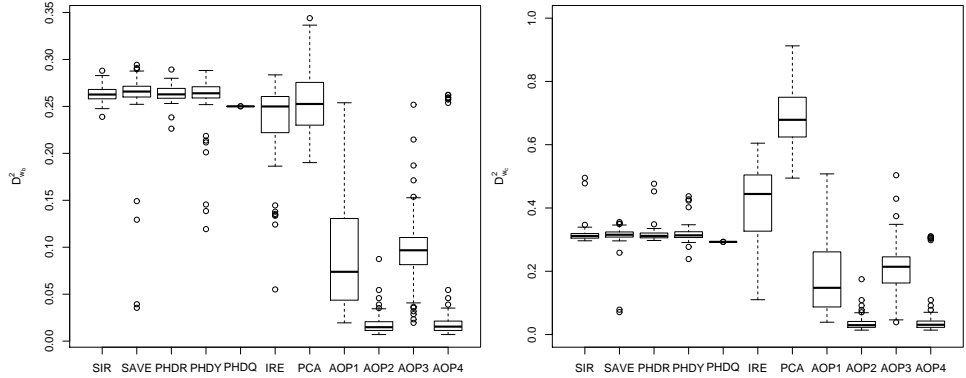}
    \caption{ Boxplots of the observed $D_{w_b}^2$ (left panel) and $D_{w_c}^2$ (right panel) distances between the true and estimated  projection matrices
    when the observations come from the model M1 with $k=2$.}
    \label{fig: avproj_distances_M2} 
\end{figure}

\begin{figure}[tbph]
    \centering
    \includegraphics[height = 5.5cm]{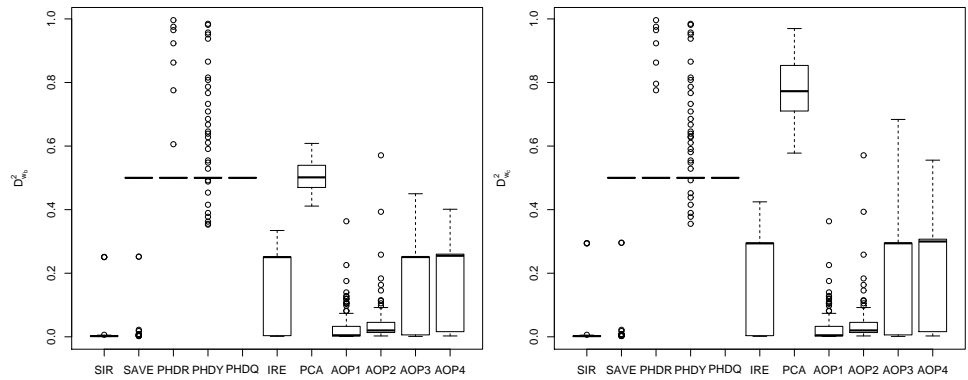}
    \caption{ Boxplots of the observed $D_{w_b}^2$ (left panel) and $D_{w_c}^2$ (right panel) distances between the true and estimated  projection matrices
    when the observations come from the model M2 with $k=1$.}
    \label{fig: avproj_distances_M4} 
\end{figure}

\begin{figure}[tbph]
    \centering
    \includegraphics[height = 5.5cm]{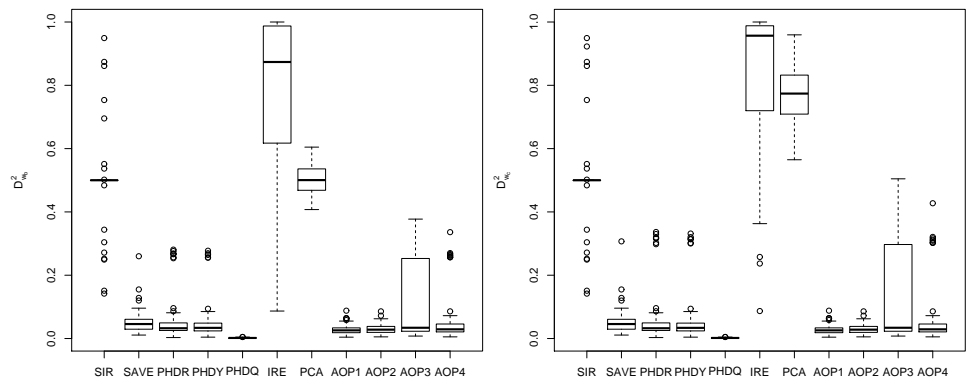}
    \caption{ Boxplots of the observed $D_{w_b}^2$ (left panel)and $D_{w_c}^2$ (right panel)distances between the true and estimated  projection matrices
    when the observations come from the model M3 with $k=1$.}
    \label{fig: avproj_distances_M6} 
\end{figure}

Some simulation results are collected in Figures~\ref{fig: avproj_distances_M2}-~\ref{fig: avproj_distances_M6}.
 The figures show the boxplots for the observed $D_{w_b}^2$ and $D_{w_c}^2$ distances between the true orthogonal projector and the projector estimates coming from different dimension reduction approaches. Consider first the behavior of the estimates in the model M1 with $k=2$ (see Figure~\ref{fig: avproj_distances_M2}). The performances of SIR, SAVE and PHD estimates seem to be very similar and they usually find only one direction. (For example, SIR finds only the component with linear dependence, and SAVE only the component of quadratic dependence.)  The same seems to be true with IRE but with more varying estimates. Recall that, if $\mathcal{S}_{P_1}\subset \mathcal{S}_{P_2}$ and $k_1=1$ and $k_2=2$, then $\lambda=k_1/k_2=0.5$, and the average distances of SIR, SAVE and PHD estimates tend to be close to
  $$D^2_{w_b}(P_1,P_2)=\frac 1{2k_1}(1-\lambda)=0.25
 \ \ \mbox{and}\ \
  D^2_{w_c}(P_1,P_2)=1-\sqrt{\lambda}=0.293,$$
 respectively. The AOP estimates then nicely pick up the two dimensions and clearly outperform other estimates. Note that there is no big difference between AOP estimates with known $k$ and AOP estimates with estimated $k$. The AOP estimate based on $w_c$ seems to be better. PCA has a poor performance as expected.

Figure~\ref{fig: avproj_distances_M4} shows the results when the observations come from the model M2. The model with linear dependence only is then of course the model where SIR is the best one. IRE also performs quite well but, for most samples,  SAVE and PHD approaches do not find any solution at all. Recall that, if $k_1=0$ and $k_2=1$ then $D^2_{w_b}(P_1,P_2)=D^2_{w_c}(P_1,P_2)=0.5$. The AOP estimates seem often to pick up the correct subspace, and there is no real difference between the
$w_b$ and $w_c$ estimates. This time, the AOP estimates with known dimension $k=1$ seem to perform better than the AOM estimates with estimated $k$.

Figure~\ref{fig: avproj_distances_M6} gives the results for the model M3 with $k=1$ and quadratic dependence.  SAVE and  PHD approaches work very well, and
SIR and IRE completely fail in this case. Again, all AOP estimates neglect the bad estimates and pick up nicely the correct one direction. As in the other cases, PCA provides a random reference method with a bad performance indeed.

\section{Final comments}

Dimension reduction and subspace estimation is a topic with increasing relevance since modern datasets become larger and larger. Different approaches have different {shortcomings}  and  combining the results coming from different approaches
might give a better total overview.
In this paper, we propose a generalization of the Crone and Crosby distance for the orthogonal projectors, a weighted distance that allows to combine subspaces of different dimensions. Some natural choices of weights are considered in detail.  The performance of three weighted distances and the combining approach is illustrated via simulations which show that each of them has its own justification depending on the problem at hand. Similar to other areas of statistics, this kind of ``model averaging'' seems to be a way to combine information from competing estimates and to give a better idea of the true model at hand.

\clearpage

\appendix
\section*{APPENDIX}

\parindent0mm

\begin{lemma}\label{AuxiliaryLemma} For two $p\times p$ orthogonal projectors $P_1$ and $P_2$ with ranks $k_1$ and $k_2$, respectively,
\[
\max\{ p-k_1-k_2,0\} \le tr(P_1P_2) \le \min\{ k_1,k_2\}.
\]
\end{lemma}

{\bf Proof}\ \  First note that  $P_1=U_1U_1'$ and $P_2=U_2U_2'$ where $U_1$ has $k_1$ orthonormal columns and $U_2$ has $k_2$ orthonormal columns. Then
$tr(P_1P_2)=||U_1'U_2||^2\ge 0$. As $tr(P_1P_2)+tr(P_1(I_p-P_2))=tr(P_1)=k_1$ and $tr(P_1P_2)+tr((I_p-P_1)P_2)=tr(P_2)=k_2$ one can conclude that
$tr(P_1P_2)\le\min\{k_1,k_2\}$. Similarly, $tr(P_1(I_p-P_2))\le \min\{k_1,p-k_2 \}$ and therefore
$tr(P_1P_2)=k_1-tr(P_1(I_p-P_2))\ge k_1- \min\{k_1,p-k_2 \}=\max\{k_1+k_2-p,0\}$, and the result follows.

\medskip

Note also that the lower and upper bounds in the above lemma are fixed. The upper bound is obtained with the choices
\[
P_1=\sum_{i=1}^{k_1} e_i e_i' \ \ \mbox{and}\ \ P_2=\sum_{i=1}^{k_2} e_i e_i',
\]
and the lower bound with the choices
\[
P_1=\sum_{i=1}^{k_1} e_i e_i' \ \ \mbox{and}\ \ P_2=\sum_{i=p-k_2+1}^{p} e_i e_i',
\]
where $e_i$ is a $p$-vector with the $i$th component one  and other components zero.

\medskip

{\bf Proof of Proposition 2.1} \ \
One easily sees that
\begin{eqnarray*}
D_w^2(P_1,P_2) &=& \frac{w^2(k_1)k_1 + w^2(k_2)k_2}{2} - w(k_1)w(k_2)tr(P_1P_2) \\
&=& m(k_1,k_2) - w(k_1)w(k_2)tr(P_1P_2),
\end{eqnarray*}
and the proof follows from Lemma~\ref{AuxiliaryLemma}.

\medskip

{\bf Proof of Lemma 2.1} \ \
As shown before,
\[
D_w^2(P_i,P) = \frac{1}{2}w^2(k_i)k_i + \frac{1}{2}w^2(k)k - w(k_i)w(k)tr(P_iP).
\]
Then
\[
\sigma_w^2(P) = \frac{1}{m} \sum_{i=1}^m D_w^2(P_i,P) = \frac{1}{2m}\sum_{i=1}^m w^2(k_i)k_i + \frac{1}{2}w^2(k)k - w(k)tr(\bar P_w P).
\]
The first term in the latest sum does not depend on $P$ or $k$. Thus, $\sigma_w^2(P)$ is minimized when $w(k)tr(\bar P_w P) - \frac{1}{2}w^2(k)k$ is maximized.

 \medskip

{\bf Proof of Proposition 2.2} \ \
The AOP $P_w$ maximizes
\[
D(P) = w(k)tr(\bar P_w P) - \frac{1}{2}w^2(k)k,
\]
where $k$ is the rank of $P$. Assume first that $k>0$ is fixed and $P=VV'$ where $V$ has $k$ orthonormal columns. Then $D(P)$ is maximized as soon as  $tr(\bar P_w P)$ is maximized.  Then, as $\bar P_w = \sum_{i=1}^p \lambda_i u_iu_i'$, $tr(\bar P_w P) = tr(\bar P_w VV') = tr(V'\bar P_w V)$ is maximized if $V=(u_1,...,u_k)$, and the maximum value is $\sum_{i=1}^k \lambda_i $.  For fixed $k>0$, the maximum value of $D(P)$ is then $w(k)\sum_{i=1}^k \lambda_i - \frac{1}{2}w^2(k)k$, and $D(0)=0$.
The result follows.

\end{document}